# Low-index contrast waveguide bend based on truncated Eaton lens implemented by graded photonic crystals

S. HADI BADRI[1,*], M. M. GILARLUE[1]

[1]*Department of Electrical Engineering, Sarab Branch, Islamic Azad University, Sarab, Iran*
*\* sh.badri@iaut.ac.ir*

**Abstract:** Low-index contrast waveguides such as silica waveguides are indispensable part of passive integrated optical components. Reducing the bending radius of silica waveguides with low bend loss is vital in miniaturizing silica planar lightwave circuits. The Eaton lens is a gradient index lens that can bend parallel light rays by 90°, 180°, or 360°. We present a low-loss and compact 90° waveguide bend with a radius of 9µm designed by truncating the Eaton lens. The performance of the designed bend, implemented by graded photonic crystal, was evaluated by full-wave two-dimensional finite element method. The average bend loss in the C-band is 0.9dB while the average bend loss for the wavelength range of 1260-1675nm is 1.05dB. The proposed design strategy can be applied to other low-index contrast waveguides.

## 1. Introduction

Low-index contrast silica waveguides, due to their low propagation loss, efficient fiber-to-chip coupling, low material birefringence ($<10^{-4}$), low cost, and high temperature stability have been utilized in planar lightwave circuits (PLCs) [1]. The key drawback of the components based on silica waveguides is their large footprint, where the large radius of waveguide bend has been identified as one of the main obstacles. The radius of silica waveguide bends could rise up to a few millimeters, severely limiting the integration of optical components [2, 3]. The effect of minimizing the bending radius is more noticeable in devices consisting of many bending waveguides such as arrayed-waveguide gratings (AWGs). Without mode-conversion interfaces, silicon photonic wire AWGs [4] poorly couple to single mode fibers. Introducing local high-index contrast by air trenches with high aspect-ratio mesa structures provides locally enhanced lateral confinement. This method has been employed to reduce the bend radius of a silica waveguide with refractive index contrast of $\Delta(n_{core}/n_{cladding}-1)=6.85\%$ to 7.25µm. However, considering the space occupied by air trenches the overall footprint of the bend rises to 20µm×20µm [3, 5]. Furthermore, the narrow mesa makes these structures fragile and the side walls are prone to dust and impurities contamination. To overcome these problems, trenches filled with low-refractive index materials with the bending radius of 200µm have been reported [6]. An air hole photonic crystal (PhC) structure occupying a large area of 27µm×27µm has also been utilized to reflect the optical wave in the desired direction. This silica waveguide bend provides a 3dB bandwidth of about 185nm [7]. Multilayer air trench as corner mirror has also been introduced as waveguide bending mechanism with about 8µm×10µm footprint and bending loss lower than 0.45dB in a bandwidth of 500nm [8]. Sharp bending radius of 2µm based on gradient refractive index (GRIN) structures in the core of the

1µm-width waveguide with index contrast of ∆=50% has been reported [9]. Variety of high-index contrast silicon waveguide bends with small radius have also been studied. Inverse designed bends with 2.6µm×2.6µm footprint, 1dB loss, and a 40nm bandwidth have been proposed [10]. Mode converter assisted multimode waveguide bends with a radius of 5µm, low loss of 0.2dB and a bandwidth of 100nm has been introduced [11]. Bending loss could be efficiently minimized by introducing the optimum ratio of clothoid curves in the bend. Bend radius of 4µm with loss of 0.002dB has been reported by this method [12]. Multimode waveguide bend with a radius of 78µm has been designed by transformation optics [13]. However, the silicon waveguides typically suffer from scattering losses because of the sidewall's roughness, and poor coupling to single-mode fiber. To achieve lower scattering losses, waveguide index contrast can be reduced. However, bend radiation loss is very high in low-index contrast waveguides [14, 15].

In this paper, we discuss an alternative approach to design a compact bend based on the Eaton lens for low-contrast waveguides. Recently, interesting applications have been introduced for GRIN lenses such as Luneburg [16, 17], Maxwell's fish-eye [18, 19], and Eaton [20] lenses. The Eaton lens can change the light wave's trajectory by 90°, 180°, or 360°. The refractive index of the Eaton lens that bends light by 90° is calculated by [21]

$$n_{lens}^2 = \frac{R_{lens}}{n_{lens}r} + \sqrt{\left(\frac{R_{lens}}{n_{lens}r}\right)^2 - 1} \qquad (1)$$

where $R_{lens}$ is the radius of the lens and $r$ is the radial distance from the center of the lens. The refractive index of the lens, $n_{lens}$, ranges from unity at its edge to infinity at the center of the lens. The radially-symmetric refractive index profile of the lens is calculated numerically. We have truncated the Eaton lens to improve the bend performance and also reduce the footprint slightly. To design a broadband waveguide bend, the effective medium theory was adopted and implemented by all-dielectric graded photonic crystal (GPC). The lens is realized by changing the filling ratio of silicon rods in the silica background.

## 2. Waveguide bend design

Waveguide bends are intrinsically lossy since the mode profile shifts towards the outer edge of the waveguide. High-index contrast waveguides such as silicon waveguides confine the optical wave well within the waveguide bend. Therefore, bend loss is relatively low in silicon waveguide bend with small radius e.g. 5µm [22]. However, silica waveguides don't have this advantage.

In this paper, a waveguide bend is designed for a doped silica waveguide with core width of $w$=4µm and refractive index of $n_{core}$=1.50. The core is surrounded by silica with refractive index of $n_{clad}$=1.46 at a wavelength of 1550nm. Fig. 1(a) illustrates the ray's trajectories in the Eaton lens with the radius of 7.5µm. Our goal is to intersect a waveguide with the Eaton lens and consequently bent the light rays with the help of the lens. To ensure the impedance match between the edge of the lens and the core of the waveguide, the calculated $n_{lens}$ is multiplied by $n_{core}$. In ray-tracing calculations, a linear array of point sources covers a width of the waveguide core ($w$). The rays that correspond to the boundaries of the waveguide core are bolded. However, the waveguides and the refractive index outside of the lens are not shown in Fig. 1(a). To avoid any reflection from the edge of the lens, the refractive index outside of the lens is considered to be $n_{core}$. It should be noted that the refractive index of the Eaton lens diverges to infinity at the center, however, we have limited the maximum refractive index in Fig 1(a) in order to present a more distinguishable refractive index profile for the reader. In addition, in this case the singularity at the center of the Eaton lens has very limited effect on the performance of the lens.

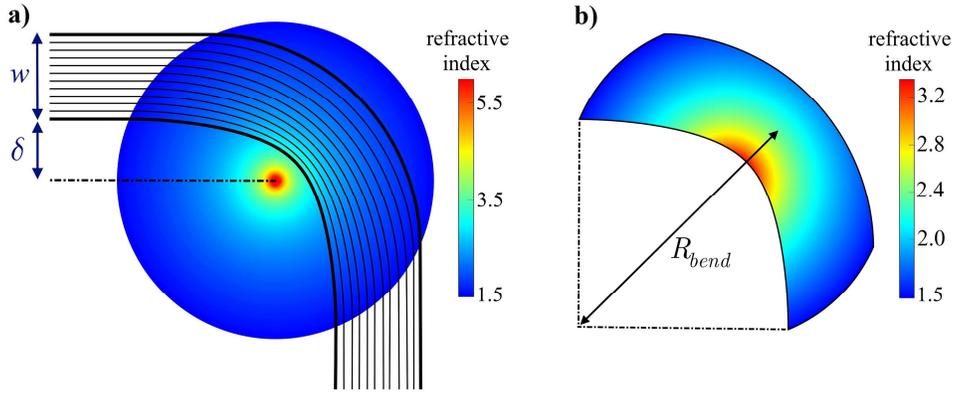

Fig. 1. a) The rays emitted from the array of point sources in a width of $w$ are bent with the Eaton lens with a radius of 7.5μm. The inner and outer rays that correspond to the boundaries of the waveguide core are bolded. The distance from the inner boundary to the center of the lens is denoted by $\delta$. b) The Eaton lens is truncated based on the inner and outer boundary rays. The effective bending radius is $R_{bend} = 9\mu m$.

Since the trajectory of rays are limited inside the bolded boundaries of Fig. 1(a), we have truncated the Eaton lens to within these boundaries. The trajectory of the bolded rays was numerically calculated and used as inner and outer boundaries of the truncated Eaton lens geometry. The geometry of the truncated Eaton lens and its refractive index profile are shown in Fig. 1(b). The outer and inner boundaries of the truncated lens can be described by:

$$\begin{cases} y = -0.05617x^2 - 0.005803x + 5.596 & \text{outer boundary} \\ y = 5.455 \times 10^{-4} x^4 - 2.262 \times 10^{-5} x^3 - 0.1101 x^2 - 2.673 \times 10^{-3} x + 1.25 & \text{inner boundary} \end{cases} \quad (2)$$

where y and x are in micrometers. As shown in Fig. 2, in the above equations the y and x axes are rotated by 45°.

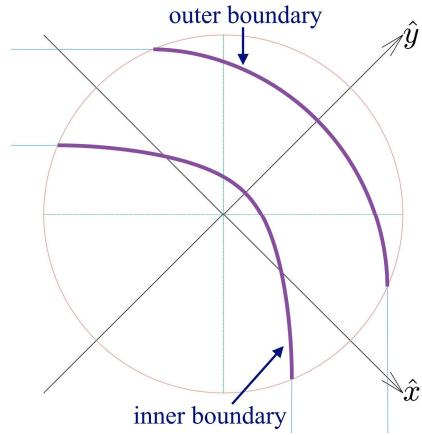

Fig. 2. The inner and outer boundaries of the truncated lens. The equation (2) describes these boundaries with respect to the y and x axes shown in this figure.

The Fig. 3(a) shows the performance of the bent waveguide without Eaton lens. In this case, the optical wave practically ignores the bent waveguide and propagates in a straight path. However, the truncated Eaton lens of Fig. 3(b) confines light wave within the bent waveguide. The bend loss of $0.33dB$ was achieved at a wavelength of $1500nm$ for the waveguide bend of Fig. 3(b). The modal field distribution in the curved section of the waveguide is different from that of the straight waveguide. As shown in Fig. 1(b), the refractive index is higher at the inner boundary of the bend while it decreases towards the outer boundary of the bend. Therefore, the wavefront moves slower at the inner side of the bend while it moves faster in the outer side of the bend. Contrary to some waveguide bends designed by transformation optic, the refractive index of the presented design at entering and exiting interfaces matches with the refractive index of the waveguide core. And the refractive index increases gradually to reduce the reflection. Therefore, the mode matching of the straight and bent waveguides is done gradually.

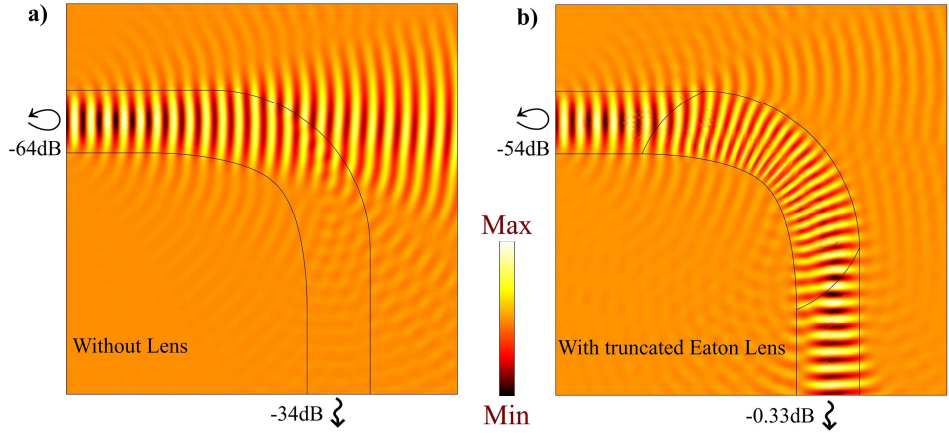

Fig. 3. The electric field distribution of the optical wave propagation though the bent waveguide a) without and b) with the truncated Eaton lens at 1550*nm*. The transmission and reflection values are shown in this figure.

The electric field distribution of the optical wave propagating through the waveguide bend based on the complete Eaton lens is shown in Fig. 4. The bend loss in this case is 1.9*dB* at 1550*nm*. By truncating the Eaton lens, we have reduced the bend loss by 1.57*dB*. This improvement is due to the fact that the truncated Eaton lens is surrounded by a lower refractive index of $n_{clad}$ and therefore, it acts as a waveguide and confines light inside it. But the complete Eaton lens does not have this advantage and it is only partially successful in guiding the optical wave towards the core of the exiting waveguide.

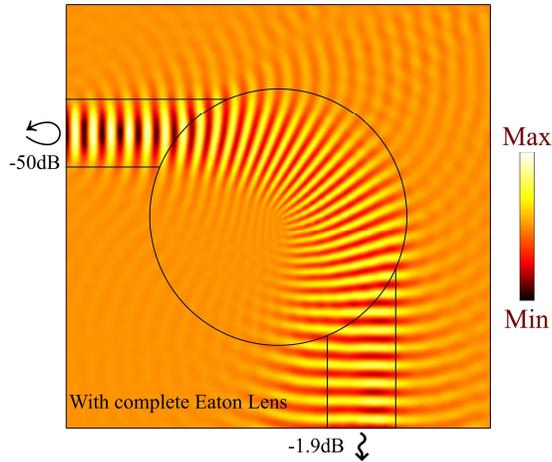

Fig. 4. The electric field distribution of the optical wave propagation though the bent waveguide with complete Eaton lens at 1550*nm*. The bend loss is higher in the complete Eaton lens compared to the truncated one.

The radius of the Eaton lens is chosen with respect to the waveguide core width in a manner that the final refractive index distribution of the truncated lens has a feasible range. Moreover, the position of the straight waveguide with respect to the center of the lens is important. The distance from the center of the lens to the inner side of the waveguide core is denoted by $\delta$ in Fig. 1(a). As shown in Fig. 5, when the waveguide is placed too close to the center of the lens, i.e. smaller $\delta$, the maximum refractive index of the truncated Eaton lens exceeds the practical values implementable by typical dielectric materials. As the distance between the waveguide and the center of the lens increases, the maximum refractive index required to implement the lens decreases. The bend loss should also be considered. As shown in Fig. 5, the bend loss also changes by varying $\delta$. To achieve a feasible refractive index range and optimum bend loss, we chose $\delta = 2.9\mu m$ for our implementation.

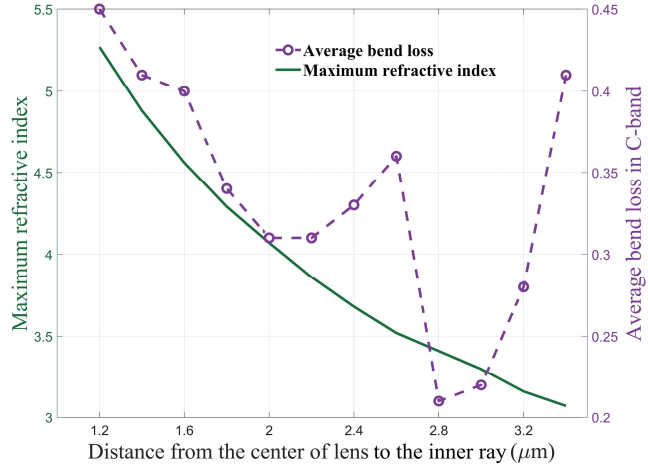

Fig. 5. The maximum refractive index required for implementing the truncated Eaton lens and its average bend loss in the C-band with respect to $\delta$ (the distance from the center of the Eaton lens to the inner ray boundary)

## 3. Graded photonic crystal implementation

Typically, metamaterials consist of resonant elements with narrow bandwidth. GPC can be regarded as a broadband all-dielectric metamaterial. The physical properties of the lens can be realized by locally varying the parameters of the unit cell of the GPC. The GPC structures have been used to implement waveguide bends [23-25]. One of the simplest methods of implementing GPC structures is to change the radius of the rod located in the middle of the unit cell, shown in Fig. 6(a). For transverse magnetic (TM) mode, where the electric field is normal to the rods, the radius of the rod in the ij-th cell is given by [19, 26]

$$r_{rod,ij} = a_{GPC}\sqrt{\frac{(n_{eff,ij}^2 - n_{host}^2)}{\pi(n_{rod}^2 - n_{host}^2)}} \quad (3)$$

where $a_{GPC}$ is the lattice constant of the GPC structure, $n_{eff,ij}$ is the effective refractive index of the ij-th cell. $n_{host}$ and $n_{rod}$ are the refractive indices of the host and rod, respectively. Minimizing the reflection at the interface of the waveguide and the lens is accomplished by matching the refractive indices of the waveguide's core and the lens's edge. Consequently, the refractive index of the lens changes gradually from $n_{core}$ at its edge to its maximum value at the middle of the bend. Accordingly, the host material in the lens should be the same as the core material of the waveguide, $n_{host}=n_{core}=1.50$. In our implementation, the rod material is chosen as silicon,

$n_{rod}=n_{Si}=3.45$. The GPC-implementation of the truncated Eaton lens of Fig. 1(b) with $a_{GPC}=183nm$ is shown in Fig. 6(b). The host material is not shown in this figure.

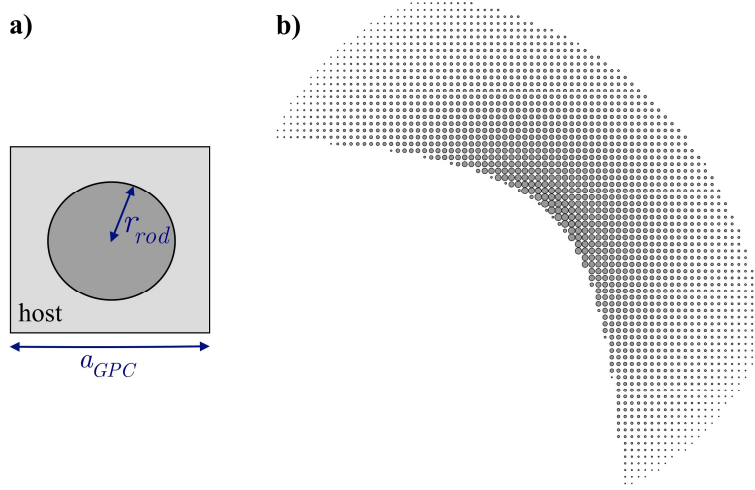

Fig. 6. a) The unit cell of the GPC structure. The radius of the rod located in the middle of the square cell with side length of $a_{GPC}$ is calculated by equation (3). b) The GPC implementation of the waveguide bend with $a_{GPC}=183nm$.

## 4. Results and discussion

The 2D finite element method (FEM) was used to evaluate the performance of the designed waveguide bend. In this paper, we focus on designing a bent silica waveguide with minimum bending radius. We have designed a bent waveguide with the effective radius of 9μm in the previous section. In our method, the bending radius increases as the width and refractive index of the waveguide's core increases. Designing a bent waveguide becomes less challenging and the bend loss decreases as the bending radius increases. The design procedure described in this paper can be applied to design bends with different radii for waveguides with different core width and refractive index. However, it should be mentioned that as the refractive index of the waveguide's core increases, the radius of the bent waveguide increases considerably. The electric field distribution of the propagating optical wave at 1550nm is illustrated in Fig. 7 for the waveguide bend implemented by GPC structure with $a_{GPC}=183nm$. The bend loss of 0.9dB and return loss of -44dB was achieved at 1550nm. The return and bend losses of the waveguide bend implemented by GPC structure with $a_{GPC}=183nm$ are shown in Fig. 8 for the optical communication bands. The average bend loss of 1.05dB was achieved for the 1260-1675nm range. It should be noted that some part of the loss is because of limited optical confinement of the low-index contrast silica waveguide. This leakage loss was not considered in our bend loss calculations. Another loss factor is the fact that the designed bend is sensitive to the continuous refractive index gradient of the lens while the GPC structure has a stepped profile.

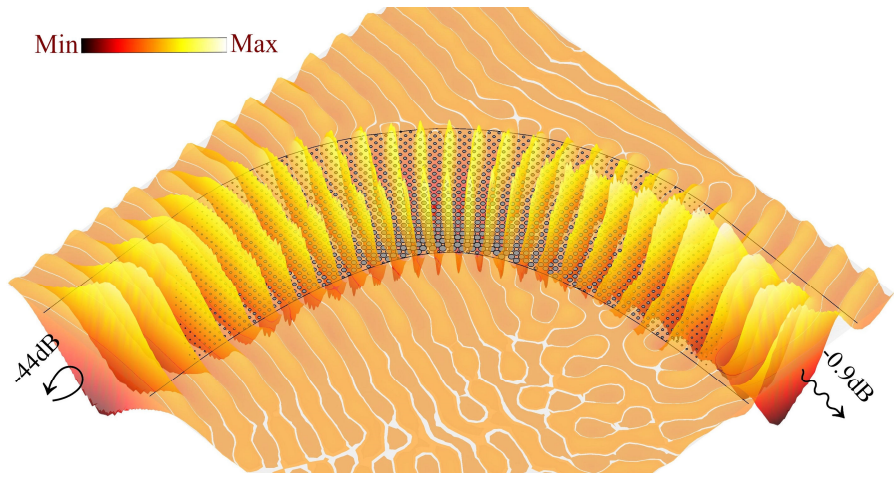

Fig. 7. The electric field distribution of the optical wave propagation though the bent waveguide implemented by GPC with $a_{GPC} = 183nm$ at 1550nm.

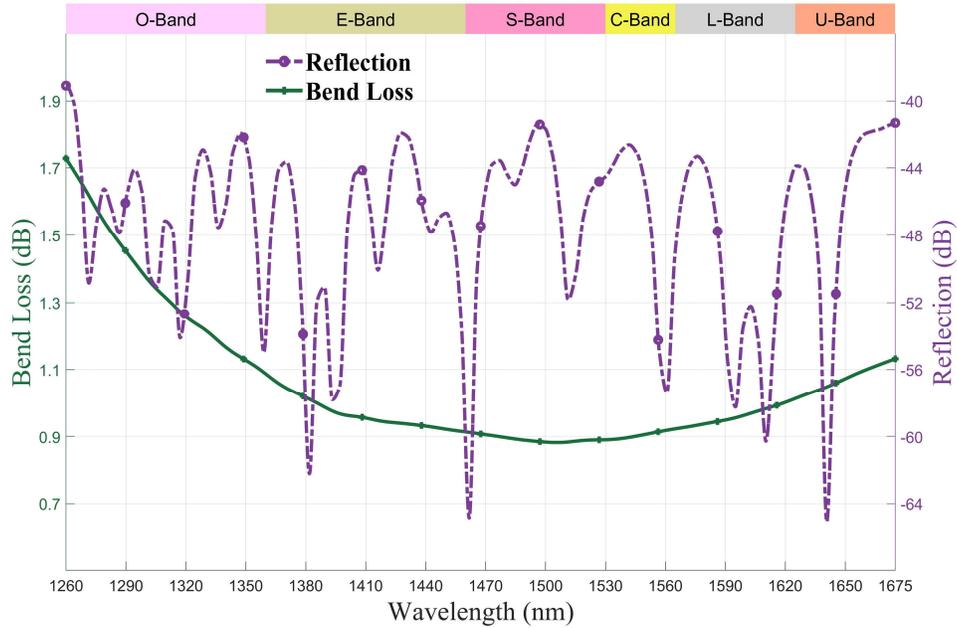

Fig. 8. The bend loss and reflection of the GPC-based bent waveguide of Fig. 7.

The Fig. 9 demonstrates the performance of the waveguide bend with the ideal truncated Eaton lens and three GPC implementations of the lens with different lattice constants of 183, 200, and 231*nm*. Obviously, as the lattice constant of the GPC structure decreases its performance approaches to the performance of the ideal lens. And the bend loss increases by increasing the lattice constant of the GPC structure. However, even with $a_{GPC} = 231nm$, the bend loss is lower than 3*dB* for 1515-1675*nm*. For low wavelengths, the optical wave does not perceive the GPC structure with $a_{GPC} = 231nm$ as an effective medium. The average bend loss for the structure with $a_{GPC} = 200nm$ is 1.73*dB*. While the average bend loss for the GPC structure with lattice constant of 183*nm* is 1.05*dB* in the 1260-1675*nm* range. An important factor that indicates the feasibility of the GPC structures is the diameter of the smallest rod. The diameter of the smallest rod in these implementations depends on the GPC structure's lattice constant. As the lattice constant decreases the diameter of the smallest rod decreases. Recent technological advances have enabled us to fabricate vertical silicon nanowires with diameters lower than 25nm [27, 28]. We limited the minimum diameter to 30nm in our implementations to slightly alleviate the fabrication constraints.

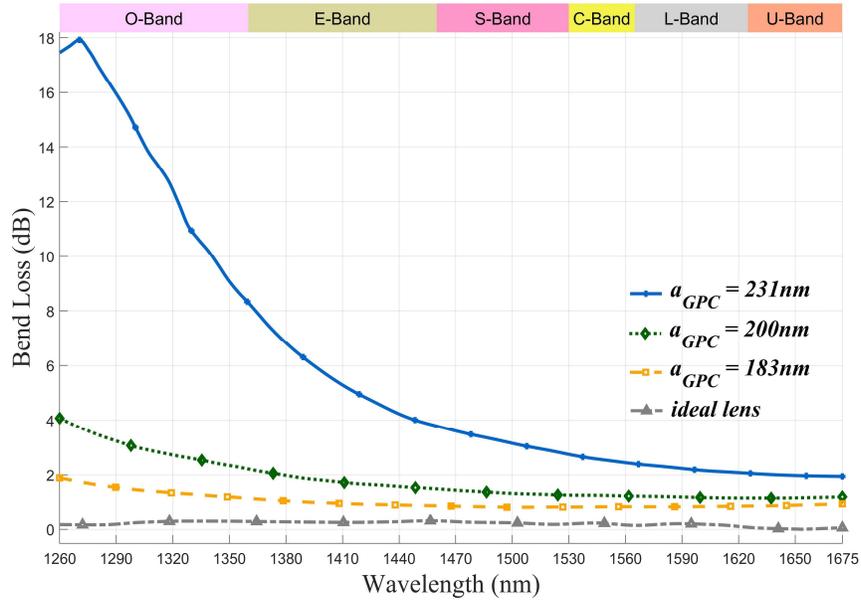

Fig. 9. Comparison of waveguide bend with the ideal truncated Eaton lens and its GPC implementation with three different lattice constants.

Bending mechanisms of the silica waveguide bends are compared in Table 1. The Bend loss, bandwidth, index contrast, and footprint of the references are summarized in this table. The reference [9] has the smallest footprint. However, the index contrast used in its design is very high compared to other works. The bend losses in references [3,6,7] are lowest, but they have limited bandwidth compared to our work. The reference [8] has a comparable bandwidth and refractive index contrast with our work, but air trenches are fragile and prone to contaminations. The presented work has not the most desirable attributes, but its footprint is comparatively low and covers the entire optical communication bands with modest bend loss.

Table 1. Comparison of silica waveguide bends

| Ref. | Bending mechanism | Δ | Bend loss(dB) | Bandwidth(nm) | Footprint(μm$^2$) |
|---|---|---|---|---|---|
| [3] | Air trench | 6.85% | 0.053 | 1530_1570 | 20×20 |
| [6] | Filled trench | 1.19% | 0.1 | 1547_1555 | 200×200 |
| [7] | PhC mirror | 0.76% | 0.17 | 1540-1565 | 27×27 |
| [8] | Air-trench mirror | 2.39% | 0.35 | 1300-1800 | 8×10 |
| [9] | GRIN structure | 50% | 0.96 | - | 2.5×2.5 |
| our work | Eaton lens | 2.74% | 1.05 | 1260-1675 | 14×14 |

## 5. Conclusion

Waveguide bends are essential building blocks for changing the propagation direction of the light wave and connecting various integrated optical components. Designing sharp bends in high-index contrast waveguides such as silicon waveguides is easier due to strong confinement of light. On the other hand, designing low-index contrast waveguide bends could be challenging. Alleviating the obstacle of large footprint introduced by low-index contrast waveguide bends is essential in integration of PLC components. In this paper, reducing the bending radius of low-index contrast waveguides with low bend-loss is achieved by utilizing the Eaton lens. 2D FEM simulations show that the GPC-based truncated Eaton lens can reduce the bend loss from 34dB to 1.05dB for a 9μm bend radius. The proposed design procedure can be used to implement bends with different radii for low-index contrast waveguides.